\documentclass[a4paper]{article}
\usepackage{amsmath}
\usepackage{amsfonts}
\usepackage{amssymb}

\title{DOMAIN WALL IN NAMBU-JONA-LASINIO MODEL}
\author{Sergii Kutnii\\Bogolyubov Institute for theoretical Physics}

\begin{document}

\maketitle

\begin{abstract}
 The approximation for the NJL gap equation that was developed in our previous paper allows 
us to investigate vacuum inhomogeneities in the mean field approach. The simplest case
of a domain wall is studied thoroughly. The Jackiw-Rebbi problem which arises in treatment of the
fermions' interaction with the
domain wall is solved explicitly and the entire spectrum of bound states is described.
It is estabilished that under our assumptions no bound states exist at the domain wall,
however higher-order corrections may make them appear.
Thus existence of the quark bound states at domain walls may be a good test for
subtle features of QCD.
\end{abstract}

\section{Introduction}

Deriving quark confinement from QCD is one of the most intriguing problems 
in modern quantum field theory. Nonlinearity of theory's Yang-Mills equations is a huge obstacle.
 It's well known that at the low energies spontaneous gauge symmetry breaking occurs. 
However the gluon propagator structure is unknown which makes it difficult to build any plausible approximations.

Nambu-Jona-Lasinio model first proposed long before QCD was developed\cite{NJL} 
is believed to be one of such approximations. In one of our previous works\cite{Me1} 
we've proposed a derivation of it from QCD in the mean field approach to the latter 
developed by Kondo\cite{Kondo}
It's notable that in our approach NJL model is obtained naturally 
together with a cutoff parameter which is equal to the effective gluon mass.

\begin{equation}
\Lambda = m_\mathcal{A}
\end{equation}

Another nontrivial feature of our approach is that the interaction constant 
of the effective NJL lagrangian depends on the cutoff parameter too:

\begin{eqnarray}
&&\mathcal{L}_q = \bar{\psi}_{i}\left(i\hat{\partial} - m_0\right)\psi_i
-\nonumber\\&& + \frac{1}{4\left(\rho^2 - 1\right)\Lambda^2}
\left[\bar{\psi}_i\psi_i\bar{\psi}_k\psi_k - \bar{\psi}_i\gamma^5\psi_i\bar{\psi}_k\gamma^5\psi_k\right. 
\nonumber\\ &&- \left. 
\frac{N + 2}{2N}\bar{\psi}_{i}\gamma_{\mu}\psi_{i}\bar{\psi}_{k}\gamma^{\mu}\psi_{k} - 
\frac{1}{2}\bar{\psi}_{i}\gamma^5\gamma_{\mu}\psi_{i}\bar{\psi}_{k}\gamma^5\gamma^{\mu}\psi_{k}\right]
\label{eq:NJLagrangian}
\end{eqnarray}

$\rho$ is a free parameter that arises in the mean field approximation to QCD 
and $N$ is the number of colors which we are keeping for the sake of possible generalisations. 

Mean field approach is also a paradigm for dealing with NJL model itself. 
It was first applied by Nambu and Jona-Lasinio in their seminal work 
where they've found the model's condensate assuming homogeneity 
i.e. that the mean field is constant over the spacetime.
Finding inhomogeneous configurations is tricky 
since one needs to solve the nonlinear integral gap equation. 
Little progress has been made on this path so far 
with the works of M. Thies et al. \cite{Thies} being a notable exception.

We've also followed this path in our last work\cite{Me2011}. There, we've 
developed a differential approximation for the gap equation based on the study 
of divergencies of the mean field action. The equation turned out to be 
Landau-Ginzburg-like containing derivatives up to second and nonlinearities
up to third order. But here the superconductivity analogy ends 
since the equation is much more complicated than plain 
Ginzburg-Landau. Even the mean field itself is much more complex being 
a matrix variable:

\begin{equation}\widehat{\Omega} = \xi + \eta\gamma^5 + \widehat{v} + 
\gamma^5\widehat{w}.\label{eq:defomega}\end{equation}

However we've found that this equation has a simple scalar sector. In other words\
it allows to put $\widehat{\Omega} = \xi; \xi \in \mathbb{R}$. The equation 
then gets much more simple and if the NJL model fermions are initially massless
it gets reduced to the well-known $\phi^4$ case:

\begin{equation}
\Box\xi(x) -K\xi(x) + 2\xi^3(x) = 0 ,
\label{eq:phi4}
\end{equation}
where K is a constant that depends on the model's parameters and is quadratic in 
the cutoff $K \sim \Lambda^2$.

This equation has the famous one-dimensional kink solution
\begin{equation}
 \xi\left(x^\mu\right) = \mu\tanh\left[\lambda{}z\right]; z\equiv{x^3}
\label{eq:kink}
\end{equation}
which can be interpreted as a domain wall between two vacua. It's worth noting
that the exact gap equation of NJL model has two solutions with opposite signs 
in the homogeneous case so the domain wall configuration is not an artifact of our
approximation.

The mean field approach to NJL model suggests that after the mean field equations
have been solved we should substitute the solution in the Dirac equation to find
the cortresponding fermionic states:
\begin{equation}
 \left[i\widehat{\partial} - \widehat{\Omega}(x^\mu)\right]\psi = 0
\label{eq:fermion}
\end{equation}
If we substitute the kink solution (\ref{eq:kink}) for $\widehat{\Omega}$
we get the well-known Jackiw-Rebbi problem \cite{JaRe}. Jackiw and Rebbi investigated
the interaction of Dirac fermions with a $\phi^4$ scalar field kink and found 
the zero-energy localized fermionic state.

Our goal here is to pick up the task of solving (\ref{eq:fermion}) completely
filling the gap in their work. It's also worth noting that while the original
Jackiw-Rebbi case of fermions interacting with a real scalar field is somewhat
artificial our NJL background gives this problem an interesting 
physical interpretation.

\section{Jackiw-Rebbi problem in the context of NJL model}

So our task is to solve the equation
\begin{equation}
  \left[i\widehat{\partial} - \mu\tanh\left(\lambda{}z\right)\right]\psi = 0
\end{equation}
But let us first find the meaning of the kink parameters in the context of NJL model.

In \cite{Me2011} we've found that the exact form of the $\phi^4$ equation
(\ref{eq:phi4}) is
\begin{equation}
\Box\xi(x) - \left[2 - G(N,\rho)\right]Z(\Lambda, M)\xi(x) + 2\xi^3(x) = 0 
\label{eq:phi4_exact}
\end{equation}
where $$G(N,\rho) = \frac{32\pi^2\left(\rho^2 - 1\right)}{N}$$ 
and $$Z(\Lambda, M) = \frac{\Lambda^2}{\ln\left(\frac{\Lambda^2}{M^2} + 1\right).}$$

$\Lambda$ is the QCD cutoff parameter and $\rho > 1$ is a free parameter that arises
in the mean field approach to QCD. The meaning of $\rho$ is unclear by now and 
revealing it can be a task for further investigation.

$M$ is found from the homogeneous gap equation

\begin{equation}
 \frac{N}{16\pi^2\left(\rho^2 - 1\right)}\left[1 - 
M^2\frac{\ln\left(\frac{\Lambda^2}{M^2} + 1\right)}{\Lambda^2}\right] = 1
\label{eq:gap}
\end{equation}

Having the homogeneous gap equation, we can notice that the $\phi^4$ equation
now becomes

\begin{equation}
 \Box\xi(x) - 2M^2\xi(x) + 2\xi^3(x) = 0
\label{eq:physeqn}
\end{equation}

An interesting case here is the large $M$ case $M \gg \Lambda$.
After expanding the logarithm and neglecting the higher order terms
the gap equation turns then into

\begin{equation}
 M^2 = \frac{N\Lambda^2}{32\pi^2\left(\rho^2 - 1\right)}
\end{equation}

and we can conclude that the large $M$ approximation is valid when 
$\rho \rightarrow 1$.

A kink solution for the equation (\ref{eq:physeqn}) is 
\begin{equation}\xi\left(x^\mu\right) = M\tanh\left(Mx^3\right)
\label{eq:specialkink}
\end{equation}

\section{Solution}
Let us now solve the Dirac equation with $\phi^4$ kink potential. 
It's enough to study the special case when the equation becomes
\begin{equation}
 \left[i\gamma^0\partial_{\tau} + i\gamma^3\partial_{z} - \mu\tanh\left(z\right)\right]\psi(\tau,z) = 0
\label{eq:initial}
\end{equation}
Eigenstates that depend on other spatial coordinates can be constructed
from its eigenstates by the means of Lorentz transformations. The "wideness" 
parameter of a kink can be absorbed into its "height" by scaling the spacetime
variables appropriately:
\begin{eqnarray}
 \tau &=& \lambda{x}^0 \nonumber\\
z &=& \lambda{x^3}\nonumber\\
\label{eq:scaling}
\end{eqnarray}
The next step is putting $$\psi = e^{-iE\tau}
\left[\begin{array}{c}\boldsymbol{\phi}(z)\\
\boldsymbol{\chi}(z)\end{array}\right]$$
By choosing the appropriate Dirac matrix representation
we reduce the equation to the following system:
\begin{eqnarray}
 i\sigma_z\frac{d\boldsymbol\chi}{dz} + \left[E - \mu\tanh(z)\right]\boldsymbol\phi&=&0\nonumber\\
i\sigma_z\frac{d\boldsymbol\phi}{dz} + \left[E + \mu\tanh(z)\right]\boldsymbol\chi&=&0
\end{eqnarray}
Then we make an ansatz 
\begin{eqnarray}\boldsymbol\phi(z) = \phi(z)\left|\uparrow\right.\rangle\nonumber\\
\boldsymbol\chi(z) = \chi(z)\left|\uparrow\right.\rangle
\label{eq:spinup}
\end{eqnarray}
where
$\sigma_z\left|\uparrow\right.\rangle = \left|\uparrow\right.\rangle$ (we might have chosen
the other eigenvector of the Pauli matrix which wouldn't change the picture much).

The resulting system of equations for $\phi,\chi$ can be written in the following matrix form:
\begin{equation}
 \frac{d}{dz}\left[\begin{array}{c}\phi\\\chi\end{array}\right] =
\left[iE\sigma_x - \mu\tanh(z)\sigma_y\right]\left[\begin{array}{c}\phi\\\chi\end{array}\right]
\label{eq:sigmaeq}
\end{equation}
Now let us define
\begin{eqnarray}
 \left|+\right.\rangle&=&\frac{1}{\sqrt{2}}
\left[\begin{array}{c}1\\i\end{array}\right]\nonumber\\
\left|-\right.\rangle&=&\frac{1}{\sqrt{2}}
\left[\begin{array}{c}1\\-i\end{array}\right]
\end{eqnarray}
These vectors satisfy the following identities:
\begin{eqnarray}
 \sigma_y\left|+\right.\rangle&=&\left|+\right.\rangle\nonumber\\
\sigma_y\left|-\right.\rangle&=&-\left|-\right.\rangle\nonumber\\
\sigma_x\left|+\right.\rangle&=&i\left|-\right.\rangle\nonumber\\
\sigma_x\left|-\right.\rangle&=&-i\left|+\right.\rangle
\end{eqnarray}
We can now make the following ansatz:
\begin{equation}
 \left[\begin{array}{c}\phi\\\chi\end{array}\right] = 
p(z)\cosh^{-\mu}(z)\left|+\right.\rangle + q(z)\cosh^\mu(z)\left|-\right.\rangle
\end{equation}
which leads us to

\begin{equation}
 \frac{d}{dz}\left[\begin{array}{c}p(x)\\q(x)\end{array}\right] = 
\left[\begin{array}{cc}0 & E\cosh^{2\mu}(z)\\
 -E\cosh^{-2\mu}(z) & 0\end{array}\right]
\left[\begin{array}{c}p(x)\\q(x)\end{array}\right].
\end{equation}
Then we exclude $q(z)$ from this system of equations and obtain the following:
\begin{eqnarray}
&&q(z) = \frac{\cosh^{-2\mu}(z)}{E}\frac{dp}{dz}\nonumber\\
&&\frac{d^2p}{dz^2} - 2\mu\tanh(z)\frac{dp}{dz} + E^2p(z) = 0 
\label{eq:excludeq}
\end{eqnarray}
If $E=0$ $p$ and $q$ decouple from each other and we can just put $p = 1, q=0$; the other linearly independent solution diverges at the infinities and we don't take it into account.

Finally, we put $\zeta \equiv \sinh(z)$ and the last equation turns into
\begin{equation}
\left(1 + \zeta^2\right)\frac{d^2p}{d\zeta^2} + 
\left(1 - 2\mu\right)\zeta\frac{dp}{d\zeta} + E^2p(\zeta) = 0
\label{eq:imgegenbauer}
\end{equation}
This equation falls into the hypergeometric class, furthermore, it can be easily proven that there's
a series of polynomial solutions 
that are orthogonal at $(-\infty,\infty)$ with the measure 
\begin{equation}
W(\zeta) = \left(1 + \zeta^2\right)^{-\mu - \frac{1}{2}}.
\label{eq:measure}
\end{equation}
Their spectrum is 
\begin{equation}
 E_n^2 = n\left(2\mu - n\right)
\label{eq:spectrum}
\end{equation}
It can be obtained by studying the asymptotics of the equation and taking into account that the leading order should disappear.
These solutions with $|n| < \mu$ correspond to the bound states.
The Rodrigues formula for these polynomials is 
\begin{equation}
p_n(\zeta) = \left(1 + \zeta^2\right)^{\mu + \frac{1}{2}}\frac{d^n}{d\zeta^n}\left(1 + \zeta^2\right)^{-\mu - \frac{1}{2}}.
\label{eq:defpoly}
\end{equation}
One can note that the polynomials are very similar 
to the well-known 
Gegenbauer polynomials.  
See the appendix for more details.

\section{Discussion}

Now, we have established that the equation (\ref{eq:initial}) has a series of bound-state solutions
with the spectrum (\ref{eq:spectrum}). However the $E = 0$ solution that does not have its 
opposite-energy counterpart should be considered the fermionic vacuum and the symmetry 
between opposite energy solutions corresponds to fermion-antifermion dualism.
Therefore, we can conclude that the actual bound states exist only if $\mu > 1$.

Let's now recall the transformation formulae (\ref{eq:scaling}) 
and the actual parameters of the NJL kink (\ref{eq:specialkink}). 
After doing this we can make a very interesting observation that under our approximation 
exactly the "borderline" $\mu = 1$ case takes place. Furthermore, this does not depend on the 
NJL condensate and thus it doesn't depend on the cutoff!

Let us now look once more at the scaling formulae (\ref{eq:scaling}). The one of interest is
$$
\mu = \frac{H}{\lambda}
$$
where $H$ is the height of the kink and $\frac{1}{\lambda}$ is its wideness. 
Therefore we can conclude that the wider is the kink the deeper the effective potential well becomes.
So we may expect that if some subtle effects of QCD effectively widen the domain wall
then binding of fermions to it can take place.

So existence of fermionic bound states at a domain wall may be a good test for those subtle effects.

\section{Appendix. Orthogonality}

The bound states' wavefunctions should be orthogonal:
\begin{equation}
\int{}d^3x\psi_n^+(x,y,z)\psi_m(x,y,z) = N(n)\delta_{nm}.
\end{equation}

In our case the states are actually "semibound" since the fermion motion 
is restricted along the z-axis along. Thus the normalization constant will be infinite. 
So we actually should prove that this scalar product reduces to

\begin{equation}
\int{}d^3x\psi_n^+(x,y,z)\psi_m(x,y,z) =\int{dxdy}\int{}dz\psi_n^+(z)\psi_m(z)
\end{equation}
and the z-integral is actually finite and $\int{}dz\psi_n^+(z)\psi_m(z) = C_n\delta_{nm}$ 
for the wavefunctions that don't depend on $x,y$ that were studied above.

To achieve this we should first reconstruct the complete Dirac spinors for the 
(\ref{eq:defpoly}) solutions. Taking into account the transformations from section 3 we obtain
\begin{equation}
 \psi_{n\uparrow}(z) = \frac{\cosh^{-\mu}(z)}{\sqrt{2}}\left[
\begin{array}{c}
\left(p_n(\sinh(z)) +\frac{1}{E_n}\frac{dp_n(\sinh(z))}{dz})\right)|\uparrow\rangle\\
i\left(p_n(\sinh(z)) -\frac{1}{E_n}\frac{dp_n(\sinh(z))}{dz})\right)|\uparrow\rangle
\end{array}\right]
\end{equation}

Thus the normalization condition becomes

\begin{eqnarray}
 &&\langle{n\uparrow}|{m\uparrow}\rangle = \int{dxdy}\int{dz}\cosh^{-2\mu}(z)\times\nonumber\\
&&\times
\left[p_n(\sinh(z))p_m(\sinh(z)) +\frac{1}{E_nE_m}\frac{dp_n(\sinh(z))}{dz}\frac{dp_m(\sinh(z))}{dz}\right]
\end{eqnarray}

However for the bound states we have, by substituting again $\sinh(z) = \zeta$

\begin{eqnarray}
&&\int{dz}\cosh^{-2\mu}(z)
\left[p_n(\sinh(z))p_m(\sinh(z)) +\frac{1}{E_nE_m}\frac{dp_n(\sinh(z))}{dz}\frac{dp_m(\sinh(z))}{dz}\right]=\nonumber\\
&& = \int\limits_{-\infty}^{+\infty}d\zeta\left(1 + \zeta^2\right)^{-\mu-\frac{1}{2}}
\left[p_n(\zeta)p_m(\zeta) + \frac{1 + \zeta^2}{E_nE_m}\frac{dp_n(\zeta)}{d\zeta}\frac{dp_m(\zeta)}{d\zeta}\right] = \nonumber\\
&&=\int\limits_{-\infty}^{+\infty}d\zeta\left(1 + \zeta^2\right)^{-\mu-\frac{1}{2}}
\left[p_n(\zeta) - 
\frac{1 - 2\mu}{E_nE_m}\zeta\frac{dp_n(\zeta)}{d\zeta} - \frac{1 + \zeta^2}{E_nE_m}\frac{d^2p_n(\zeta)}{d\zeta^2}\right]p_m(\zeta) = \nonumber\\
&& = \left(1 + \frac{E_n}{E_m}\right)\int\limits_{-\infty}^{+\infty}{d\zeta}
\left(1 + \zeta^2\right)^{-\mu-\frac{1}{2}}p_n(\zeta)p_m(\zeta)
\end{eqnarray}

Therefore we can conclude that solutions with $E_n = -E_m$ (we can make a convention 
$E_{-n} =  - E_n, p_{-n}(\zeta) \equiv p_n(\zeta)$) are orthogonal. For the rest of the cases 
scalar product reduces to $\int\limits_{-\infty}^{+\infty}{d\zeta}
\left(1 + \zeta^2\right)^{-\mu-\frac{1}{2}}p_n(\zeta)p_m(\zeta)$ and we can see that the measure is 
exactly (\ref{eq:measure}). 

Only the opposite spin case remains now uninvestigated. However orthogonality of the opposite spin solutions 
is guaranteed trivially by $\langle\uparrow|\downarrow\rangle = 0$ and the spin-down solution are
very similar to the spin-up case studied above, that's why we aren't going into studying them in detail.

\section{Acknowledgements}
Author would like to thank P.I Holod and Yu. M. Bernatska with whom he discussed the topic.

\end{document}